\documentclass{emulateapj}
\usepackage{epsfig,natbib}
\usepackage{graphicx}
\newcommand{\ms}{\mbox{m\,s$^{-1}~$}}

\newcommand{\mse}{\mbox{m\,s$^{-1}$}}

\newcommand{\mearth}{$M_\earth$~}
\newcommand{\mearthe}{$M_\earth$}
\newcommand{\rearth}{$R_\earth$~}
\newcommand{\rearthe}{$R_\earth$}

\newcommand{\msinie}{$M \sin i$}
\newcommand{\feh}{\ensuremath{[\mbox{Fe}/\mbox{H}]}}
\newcommand{\rphk}{\ensuremath{R'_{\mbox{\scriptsize HK}}}}
\newcommand{\lrphk}{\ensuremath{\log{\rphk}}}
\newcommand{\caii}{\ion{Ca}{2} H \& K}
\newcommand{\caiih}{\ion{Ca}{2} H}

\newcommand{\etaearth}{$\mathbf \eta_{\oplus} ~$}

\newcommand{\msini}{\ensuremath{M \sin i}}

\newcommand{\plaPer}{\ensuremath{9.494\pm0.005}}             
\newcommand{\plaTc}{\ensuremath{15375.01\pm0.64}}      
\newcommand{\plaK}{\ensuremath{2.75\pm0.39}}      
\newcommand{\plaMsini}{\ensuremath{8.2\pm1.2}}      
\newcommand{\plaA}{\ensuremath{0.0831\pm0.0011}}      
\newcommand{\plaNobs}{\ensuremath{96}}      
\newcommand{\plaMedUnc}{\ensuremath{0.74}}      
\newcommand{\plaJitter}{\ensuremath{1.5}}      
\newcommand{\plaRMS}{\ensuremath{2.78}}      
\newcommand{\plaChiNu}{\ensuremath{1.59}}      
\newcommand{\plaMv}{\ensuremath{6.27}}      
\newcommand{\plaBV}{\ensuremath{0.80}}      
\newcommand{\plaVmag}{\ensuremath{7.78}}      
\newcommand{\plaJmag}{\ensuremath{6.203}}      
\newcommand{\plaHmag}{\ensuremath{5.821}}      
\newcommand{\plaKmag}{\ensuremath{5.734}}      
\newcommand{\plaDist}{\ensuremath{21.1\pm0.33}}      
\newcommand{\plaFeh}{\ensuremath{-0.23\pm\mathrm{0.03}}}      
\newcommand{\plaTeff}{\ensuremath{5170\pm\mathrm{44}}}      
\newcommand{\plaVsini}{\ensuremath{0.5\pm0.5}}      
\newcommand{\plaLogg}{\ensuremath{4.63\pm0.06}}      
\newcommand{\plaLstar}{\ensuremath{\mathrm{0.30}\pm\mathrm{0.02}}}      
\newcommand{\plaMstar}{\ensuremath{\mathrm{0.78}\pm\mathrm{0.02}}}      
\newcommand{\plaRstar}{\ensuremath{\mathrm{0.68}\pm\mathrm{0.02}}}      
\newcommand{\plaLogRphkRange}{\ensuremath{-4.95} to \ensuremath{-5.00}}      
\newcommand{\plaSvalRange}{\ensuremath{0.169} to \ensuremath{0.197}}      
\newcommand{\plbPer}{\ensuremath{74.39\pm0.12}}             
\newcommand{\plbTc}{\ensuremath{15173.2\pm2.0}}      
\newcommand{\plbEcc}{\ensuremath{0.30\pm0.09}}      
\newcommand{\plbK}{\ensuremath{4.07\pm0.41}}      
\newcommand{\plbMsini}{\ensuremath{21.6\pm2.0}}      
\newcommand{\plbA}{\ensuremath{0.319\pm0.005}}      
\newcommand{\plbNobs}{\ensuremath{73}}      
\newcommand{\plbMedUnc}{\ensuremath{0.68}}      
\newcommand{\plbJitter}{\ensuremath{1.5}}      
\newcommand{\plbRMS}{\ensuremath{2.06}}      
\newcommand{\plbChiNu}{\ensuremath{1.17}}      
\newcommand{\plbMv}{\ensuremath{6.13}}      
\newcommand{\plbBV}{\ensuremath{0.78}}      
\newcommand{\plbVmag}{\ensuremath{5.73}}      
\newcommand{\plbJmag}{\ensuremath{4.112}}      
\newcommand{\plbHmag}{\ensuremath{3.582}}      
\newcommand{\plbKmag}{\ensuremath{3.501}}      
\newcommand{\plbDist}{\ensuremath{8.911\pm0.024}}      
\newcommand{\plbFeh}{\ensuremath{+0.08\pm\mathrm{0.03}}}      
\newcommand{\plbTeff}{\ensuremath{5144\pm\mathrm{50}}}      
\newcommand{\plbVsini}{\ensuremath{0.5\pm0.5}}      
\newcommand{\plbLogg}{\ensuremath{4.60\pm0.06}}      
\newcommand{\plbLstar}{\ensuremath{\mathrm{0.34}\pm\mathrm{0.02}}}      
\newcommand{\plbMstar}{\ensuremath{\mathrm{0.85}\pm\mathrm{0.02}}}      
\newcommand{\plbRstar}{\ensuremath{\mathrm{0.73}\pm\mathrm{0.02}}}      
\newcommand{\plbLogRphkRange}{\ensuremath{-4.90} to \ensuremath{-5.02}}      
\newcommand{\plbSvalRange}{\ensuremath{0.169} to \ensuremath{0.226}}      

\shortauthors{Howard {et~al.}}
\shorttitle{A Super-Earth and a Neptune-mass planet}
\begin{document}
\pagenumbering{arabic}


\title{The NASA-UC Eta-Earth Program: \\
         III. A Super-Earth orbiting HD 97658 and a Neptune-mass planet orbiting Gl 785\altaffilmark{1}}
\author{
Andrew W.\ Howard\altaffilmark{2,3}, 
John Asher Johnson\altaffilmark{4}, 
Geoffrey W.\ Marcy\altaffilmark{2}, 
Debra A.\ Fischer\altaffilmark{5}, \\
Jason T.\ Wright\altaffilmark{6}, 
Gregory W.\ Henry\altaffilmark{7},
Howard Isaacson\altaffilmark{2}, \\
Jeff A.\ Valenti\altaffilmark{8},
Jay Anderson\altaffilmark{8}, and
Nikolai E.\ Piskunov\altaffilmark{9} 
}
\altaffiltext{1}{Based on observations obtained at the W.\,M.\,Keck Observatory, 
                      which is operated jointly by the University of California and the 
                      California Institute of Technology.  Keck time has been granted by both 
                      NASA and the University of California.} 
\altaffiltext{2}{Department of Astronomy, University of California, Berkeley, CA 94720-3411, USA} 
\altaffiltext{3}{Space Sciences Laboratory, University of California, 
                        Berkeley, CA 94720-7450 USA; howard@astro.berkeley.edu}
\altaffiltext{4}{Department of Astrophysics, California Institute of Technology, MC 249-17, Pasadena, CA 91125, USA}
\altaffiltext{5}{Department of Astronomy, Yale University, New Haven, CT 06511, USA}
\altaffiltext{6}{The Pennsylvania State University, University Park, PA 16802, USA}
\altaffiltext{7}{Center of Excellence in Information Systems, Tennessee State University, 
                        3500 John A.\ Merritt Blvd., Box 9501, Nashville, TN 37209, USA}
\altaffiltext{8}{Space Telescope Science Institute, 3700 San Martin Dr., Baltimore, MD 21218, USA}
\altaffiltext{9}{Department of Astronomy and Space Physics, Uppsala University, 
                        Box 515, 751 20 Uppsala, Sweden}

\begin{abstract}
We report the discovery of planets orbiting two bright, nearby early K dwarf stars, HD\,97658 and Gl\,785.
These planets were detected by Keplerian modelling of radial velocities measured with Keck-HIRES 
for the NASA-UC Eta-Earth Survey.
HD\,97658\,b is a close-in super-Earth with minimum mass $\msini = \plaMsini$\,\mearthe,
orbital period $P = \plaPer$\,d, and an orbit that is consistent with circular.
Gl\,785\,b is a Neptune-mass planet with 
$\msini = \plbMsini$\,\mearthe,
$P = \plbPer$\,d, 
and orbital eccentricity $e = \plbEcc$.
Photometric observations with the T12 0.8\,m automatic photometric telescope at Fairborn Observatory  
show that HD\,97658 is photometrically constant at the radial velocity period to 0.09\,mmag, 
supporting the existence of the planet.
\end{abstract}

\keywords{planetary systems --- stars: individual (HD\,97658, Gl\,785) --- techniques: radial velocity}

\section{Introduction}
\label{sec:intro}

Radial velocity (RV) searches for extrasolar planets are
discovering less massive planets by taking advantage of
improved instrumental precision, higher observational cadence, and 
diagnostics to identify spurious signals.  
These discoveries include planets with minimum masses (\msinie) as low as 1.9\,\mearth \citep{Mayor09} and 
systems of multiple low-mass planets \citep{Lovis2006,Fischer08,Vogt2010}.    
To date, 15 planets with \msini\,$<$\,10\,\mearth 
and 18 planets with \msini\,$=$\,10--30\,\mearth have been 
discovered by the RV technique \citep[Exoplanet Orbit Database\footnote{http://exoplanets.org}]{Wright2010}.  
Transiting searches for extrasolar planets have detected Neptune-mass planets 
\citep{Bakos09,Hartman2010} and  
super-Earths \citep{Leger09,Charbonneau09}.  
The initial data release from the \textit{Kepler} mission shows substantially increasing planet occurrence 
with decreasing planet radius \citep{Borucki10b}.  
Using the large number of low-mass planets, we can statistically study  
planet properties, occurrence rates, and parameter correlations in ways previously 
only possible with higher mass gas-giant planets.

We recently completed an analysis of close-in planet occurrence 
for 166 G- and K-type dwarf stars in the Eta-Earth Survey \citep{Howard2010b}.  
We studied the planet detections and non-detections on a star-by-star basis, 
estimating search completeness.  
We detected increasing planet occurrence with decreasing planet mass 
over the mass range 3--1000\,\mearth for planets with orbital periods $P < 50$\,d.  
We parameterized the planet mass distribution with a power law model 
from which we extrapolated the occurrence rate of close-in Earth-mass planets, 
giving \etaearth = $23^{+16}_{-10}$\% for planets in the mass range 0.5--2.0\,\mearth with $P < 50$\,d.

Our study also addressed a key prediction of population synthesis models of 
planet formation \citep{Ida04a,Ida_Lin08_v,Mordasini09a}---the 
expected dearth of close-in, low-mass planets.  
The ``desert'' emerges in the simulations from fast migration and accelerating planet growth.
Most planets are born near or beyond the ice line and 
those that grow to a critical mass of several Earth masses  
either rapidly spiral inward to the host star 
or undergo runaway gas accretion and become massive gas giants.  
Our measurements contradict this prediction; 
we found the highest occurrence rate for planets where theory predicted a dearth, 
in the regime of 3--30\,\mearth and $P < 50$\,d.
Population synthesis models of planet formation are 
currently unable to explain the distribution of low-mass planets.

To measure the planet occurrence rate as a function of planet mass,
our study included previously detected planets, as well as unannounced ``planet candidates''  \citep{Howard2010b}.  
Including candidates was necessary to reliably estimate occurrence fractions for low-mass planets, 
even though the candidates had formal false alarm probabilities (FAPs) 
as large as 5\% at the time of our analysis (June 2010).  
Such an FAP implies that the planet is very likely to exist, 
but it's too high for the secure announcement of a definite planet detection with well-measured orbital parameters.  
Since then, we continued to intensively observe the planet candidates.  
Based on the new confirmatory data we report two of them here as bona fide planets.  
We present HD\,97658\,b, a close-in, super-Earth planet identified as ``Candidate 3'' in \cite{Howard2010b}, 
and Gl\,785\,b, a Neptune-mass planet identified as  ``Candidate 7''.

Below we describe the host stars (Section \ref{sec:props}) 
and the RV measurements (Section \ref{sec:obs}).  
We analyze these measurements with Keplerian models 
and assess the probability of spurious detections by computing 
false alarm probabilities (Sections \ref{sec:hd97658} and \ref{sec:gl785}).
We describe photometric observations of HD\,97658 and 
the limits they impose on  planetary transits (Section \ref{sec:photometry}).
We discuss the radii of these planets and a trend in the host star metallicities 
among low-mass planets (Section \ref{sec:discussion}).

\section{Stellar Properties}
\label{sec:props}


We used Spectroscopy Made Easy \citep{Valenti96}
to fit high-resolution spectra of HD\,97658 (HIP\,54906, GJ\,3651) 
and Gl\,785 (HD\,192310, HIP\,99825),
using the wavelength intervals, line data, and methodology
of \citet{Valenti05}.  We further constrained surface
gravity using Yonsei-Yale (Y$^2$) stellar structure models
\citep{Demarque04}  and revised \textit{Hipparcos} parallaxes
\citep{vanLeeuwen07}, using the iterative method of \citet{Valenti09}. 
The resulting stellar parameters listed in
Table \ref{tab:stellar_params} are effective temperature, surface gravity, iron
abundance, projected rotational velocity, mass, radius, and luminosity. 
Both stars are K dwarfs on the main sequence.

HD\,97658 lies 0.46 mag below 
the $Hipparcos$ average main sequence ($M_V$ versus $B-V$) as defined by \citet{Wright05}.  
This location is consistent with the low metallicity of \feh\ = \plaFeh.
Gl\,785 is 0.06 mag above the $Hipparcos$ average main sequence, 
consistent with its slightly super-solar metallicity of \feh\ = \plbFeh.

Measurements of the cores of the \caii\ lines of each spectrum show 
low levels of chromospheric activity, 
as quantified by the $S_{\mathrm{HK}}$ and \lrphk.
These chromospheric indices show long-term trends over the six years of measurements, 
possibly partial activity cycles, so we list ranges of activity indices in Table \ref{tab:stellar_params}.
We detect a weak correlation between individual RVs and $S_{\mathrm{HK}}$ measurements 
for HD\,97658, but not for Gl\,785.  
This correlation, with a Pearson linear correlation coefficient of $r$\,=\,$+$0.35, 
does not appear to affect the Keplerian fit of HD\,97658\,b 
because the  $S_{\mathrm{HK}}$ time series has negligible Fourier power at or near the adopted orbital period, 
even when the long-term activity trend is removed.

Following \citet{Isaacson2010}, and based on $S_{\mathrm{HK}}$, $M_V$, and $B-V$, 
we estimate an RV jitter of 1.5\,\ms for these stars.  
This empirical estimate is based on an ensemble of stars with similar characteristics 
and accounts for RV variability due to 
rotational modulation of stellar surface features, stellar pulsation, undetected planets, 
and uncorrected systematic errors in the velocity reduction \citep{Saar98,Wright05}.  
Jitter is added in quadrature to the RV measurement uncertainties for Keplerian modelling.

\begin{figure}
\epsscale{1.15}
\plotone{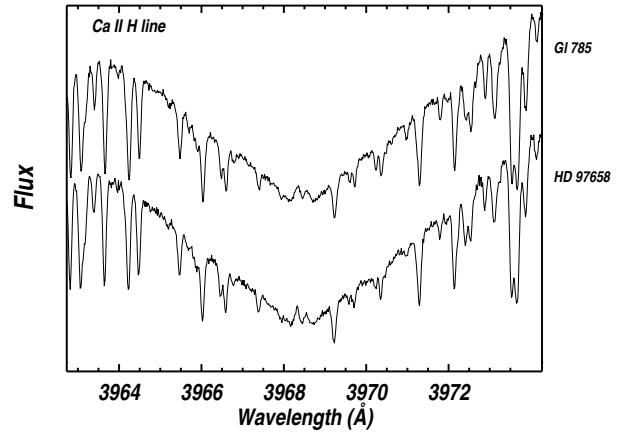}
\caption{Keck-HIRES spectra of the \caiih\ line of the early K dwarf stars HD\,97658 and Gl\,785.  
Slight line core emission near 3968\,\AA\ indicates modest chromospheric activity.}
\label{fig:caii}
\end{figure}

\section{Keck-HIRES Velocity Measurements}
\label{sec:obs}

We observed HD\,97658 and Gl\,785 with the HIRES echelle spectrometer \citep{Vogt94} 
on the 10-m Keck I telescope.  
The observations of each star span six years (2004--2010). 
All observations were made with an iodine cell mounted directly in front of the 
spectrometer entrance slit.  The dense set of molecular absorption lines imprinted 
on the stellar spectra provide a robust wavelength fiducial 
against which Doppler shifts are measured, 
as well as strong constraints on the shape of the spectrometer instrumental profile at 
the time of each observation \citep{Marcy92,Valenti95}.

We measured the Doppler shift of each star-times-iodine spectrum using a 
modelling procedure descended from \citet{Butler96b} as described in \citet{Howard2010a}.  
%
The velocity and corresponding uncertainty for each
observation is based on separate measurements for $\sim$700 spectral chunks each 2 \AA\ wide.
Once the two planets announced here emerged as candidates (about two years ago) 
we increased the nightly cadence of measurements and 
made three consecutive observations per night to reduce the 
Poisson noise from photon statistics.  
We calculate one mean velocity for multiple observations in a 2\,hr interval.

The highest RV measurement precision using Keck-HIRES has been achieved on 
chromospherically inactive late G and early K dwarfs, like the two stars presented here.  
The quietest of these stars are stable over many years at the $\sim$1.5--2.0\,\ms level
\citep{Howard09a,Howard09b,Howard2010a}; 
velocity residuals are due to astrophysical perturbations,
instrumental/systematic errors, and Poisson noise.
All of the measurements reported here were made after the HIRES CCD upgrade 
in 2004 August and do not suffer from the higher noise and systematic errors that limited the 
precision of pre-upgrade measurements to $\sim$2--3\,\ms for most stars. 

For each star we constructed a single-planet Keplerian model 
using the orbit fitting techniques described in \citet{Howard09b} 
and the partially linearized, least-squares fitting procedure described in \citet{Wright09b}.  
The Keplerian parameter uncertainties for each planet were derived using a 
Monte Carlo method \citep{Marcy05} and do not account for correlations between parameter errors.
Uncertainties in \msini\ reflect uncertainties in $M_{\star}$ and the orbital parameters.

\begin{deluxetable}{lcc}
\tabletypesize{\footnotesize}
\tablecaption{Stellar Properties of HD\,97658 and Gl\,785
\label{tab:stellar_params}}
\tablewidth{0pt}
\tablehead{
  \colhead{Parameter}   & 
  \colhead{HD\,97658} &
  \colhead{Gl\,785} 
}
\startdata
Spectral type ~~~~~~& K1\,V& K1\,V\\
$M_V$ & \plaMv & \plbMv \\
$B-V$  & \plaBV & \plbBV \\
$V$    & \plaVmag & \plbVmag \\
$J$    & \plaJmag & \plbJmag \\
$H$    & \plaHmag & \plbHmag \\
$K$    & \plaKmag & \plbKmag \\
Distance (pc) & \plaDist & \plbDist\\
$T_\mathrm{eff}$ (K) &  \plaTeff & \plbTeff \\
log\,$g$ & \plaLogg & \plbLogg \\
\feh & \plaFeh & \plbFeh \\
$v$\,sin\,$i$ (km\,s$^{-1}$) & \plbVsini &  \plaVsini \\
$L_{\star}$ ($L_{\sun}$) & \plbLstar & \plaLstar \\
$M_{\star}$ ($M_{\sun}$) & \plbMstar & \plaMstar \\
$R_{\star}$ ($R_{\sun}$) & \plbRstar & \plaRstar \\
\lrphk & \plaLogRphkRange & \plbLogRphkRange \\
$S_\mathrm{HK}$ & \plaSvalRange & \plbSvalRange \\
\enddata
\end{deluxetable}

\section{HD\,97658}
\label{sec:hd97658}

\begin{figure}
\epsscale{1.15}
\plotone{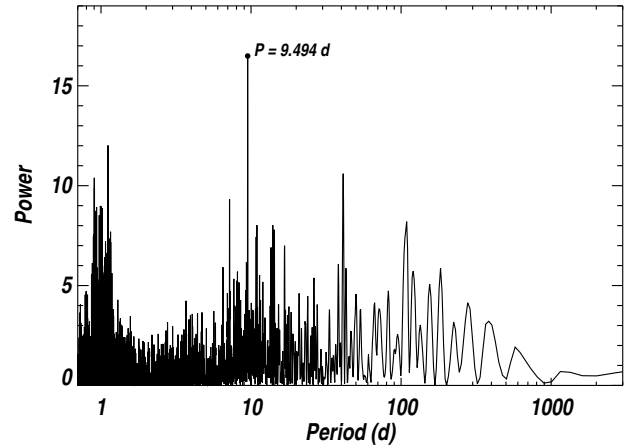}
\caption{Lomb-Scargle periodogram of RV measurements of HD\,97658.
     The tall peak near $P=9.494$\,d suggests a planet with that orbital period.
     }
\label{fig:pergram_97658}
\end{figure}

The RVs and $S_\mathrm{HK}$ values from Keck-HIRES  are listed in 
Table \ref{tab:keck_vels_hd97658}.  
Figure \ref{fig:pergram_97658} shows  a Lomb-Scargle periodogram \citep{Lomb76,Scargle82} 
of the RVs with a substantial peak at 9.494\,d.  
We used that period, as well as a wide variety of other trial periods, 
as seeds for the Keplerian fitting algorithm \citep{Wright09b}.  
Our search identified the single-planet orbital solution listed in 
Table \ref{tab:orbital_params_hd97658} as the best fit.  

We also tried fitting the RVs with an eccentric Keplerian model and found 
a best-fit solution with a nearly identical orbital period and $e = 0.17\pm0.17$, 
which is consistent with circular at the 1-$\sigma$ level.  
The detection of nonzero eccentricity with better than 95\% confidence (2-$\sigma$) 
requires approximately $e/\sigma_e > 2.45$, where $\sigma_e = \sigma/K \cdot (2/N)^{0.5}$,
$\sigma$ is the measurement uncertainty (including jitter), 
and $N$ is the number of uniformly phase-distributed observations \citep{Valenti09,Lucy71}.  
Our measurements do not meet this criterion.  
Furthermore, the eccentric model does not improve $\sqrt{\chi^2_\nu}$ from the circular model. 
We adopt the circular orbit model in Table \ref{tab:orbital_params_hd97658}.

\begin{deluxetable}{lc}
\tabletypesize{\footnotesize}
\tablecaption{Orbital Solution for HD\,97658\,b
\label{tab:orbital_params_hd97658}}
\tablenote{We adopt a circular orbital solution for this planet.}
\tablewidth{0pt}
\tablehead{
\colhead{Parameter}   & \colhead{Value} 
}
\startdata
$P$ (days)     & \plaPer \\
$T_c$ (JD -- 2,440,000) & \plaTc \\
$e\,^{a}$                     & $\equiv$0.0\\
$K$ (m\,s$^{-1}$)       & \plaK \\
$M$\,sin\,$i$ ($M_\earth$) & \plaMsini \\
$a$ (AU)                & \plaA \\
$N_\mathrm{obs}$ (binned) & \plaNobs \\
Median binned uncertainty (\mse) & \plaMedUnc \\
Assumed jitter (\mse) & \plaJitter \\
$\sigma$ (\mse) & \plaRMS \\
$\sqrt{\chi^2_\nu}$  & \plaChiNu \\
\enddata
\end{deluxetable}

\begin{figure}
\epsscale{1.15}
\plotone{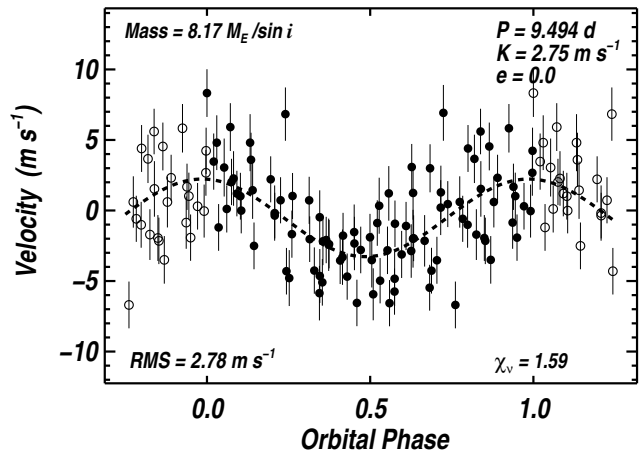}
\caption{Single-planet model for the RVs of HD\,97658, 
       as measured by Keck-HIRES.
       The dashed line shows the best-fit circular orbital solution.
       Filled circles represent phased measurements  
       while the open circles represent the same velocities wrapped one orbital phase.
       The error bars show the quadrature sum of measurement uncertainties and 1.5\,\ms jitter.}
\label{fig:phased_97658}
\end{figure}

We considered the null hypothesis---that the observed RVs are the chance arrangement 
of random velocities masquerading as a coherent signal---by calculating two false alarm probabilities (FAPs).   
Using the method described in \citet{Howard09b}, 
we computed the improvement in $\Delta\chi^2$ from a constant velocity model to a 
Keplerian model for $10^3$ scrambled data sets.  
In the first FAP test, we allowed for eccentric single-planet orbital solutions in the scrambled data sets.
We found that three scrambled data sets had a larger $\Delta\chi^2$ than the measured velocities, 
implying an FAP of $\sim$0.003 for this scenario.  
When we restricted the search for orbital solutions to circular orbits, 
none of the scrambled data sets had a larger $\Delta\chi^2$ than  measured velocities, 
implying an FAP of less than $\sim$0.001.

The rms of \plaRMS\,\ms about the single-planet model is relatively high compared 
to our adopted jitter of \plaJitter\,\ms for this chromospherically quiet K dwarf star.
This  suggests that the measured RVs are compatible with additional detectable planets.   
We computed a periodogram of the RV residuals to the single-planet fit and found  
several periods with considerable power in the range $\sim$40--200\,d.  
These peaks correspond to Doppler signals with $\sim$1.5--3\,\ms semiamplitudes.  
We considered two-planet orbital solutions with $P_b$ seeded with the best-fit value 
from the single-planet model and $P_c$ seeded with peaks in the residual periodogram.  
We allowed all orbital parameters including eccentricities to float in the two-planet 
fitting process \citep{Wright09b}.  
No two-planet solutions were found with an FAP below 5\%.  
We will continue to observe this star in search of additional planets.  

\section{Gl\,785}  
\label{sec:gl785}

\begin{figure}
\epsscale{1.15}
\plotone{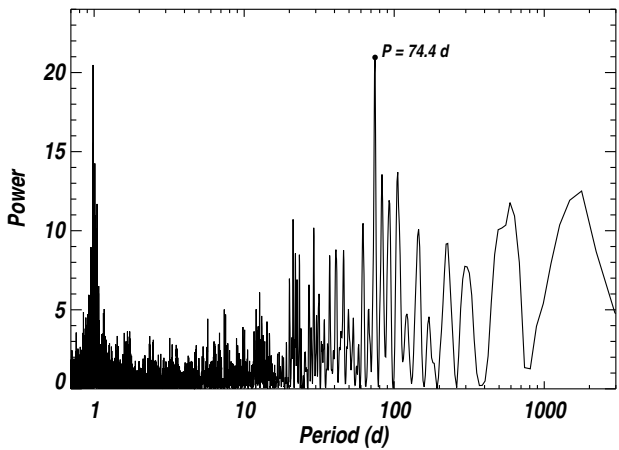}
\caption{Lomb-Scargle periodogram of RV measurements of Gl\,785.
     The tall peak near $P=74.4$\,d suggests a planet with that orbital period.}
\label{fig:pergram_192310}
\end{figure}

The RVs and $S_\mathrm{HK}$ values from the Keck-HIRES measurements of Gl\,785 are listed in 
Table \ref{tab:keck_vels_gj785}.  
Figure \ref{fig:pergram_192310} shows a Lomb-Scargle periodogram \citep{Lomb76,Scargle82} 
of the RVs with a substantial peak near 74.4\,d.  
We identify the peaks near 1.0\,d as stroboscopic aliases of the sidereal day with the 74.4\,d signal 
and other long periods \citep{Dawson2010}. 
We used 74.4\,d, as well as a wide variety of other periods, as seed periods for the 
single-planet Keplerian fitting algorithm \citep{Wright09b}.  
Our search identified the single-planet orbital solution listed in 
Table \ref{tab:orbital_params_hd97658} as the best fit.  
The orbital eccentricity of \plbEcc\ is significant at the 3-$\sigma$ level.  

\begin{deluxetable}{lc}
\tabletypesize{\footnotesize}
\tablecaption{Orbital Solution for Gl\,785\,b
\label{tab:orbital_params_gj785}}
\tablewidth{0pt}
\tablehead{
\colhead{Parameter}   & \colhead{Value} 
}
\startdata
$P$ (days)     & \plbPer \\
$T_c$ (JD -- 2,440,000) & \plbTc \\
$e$                     & \plbEcc\\
$K$ (m\,s$^{-1}$)       & \plbK \\
$M$\,sin\,$i$ ($M_\earth$) & \plbMsini \\
$a$ (AU)                & \plbA \\
$N_\mathrm{obs}$ (binned) & \plbNobs \\
Median binned uncertainty (\mse) & \plbMedUnc \\
Assumed jitter (\mse) & \plbJitter \\
$\sigma$ (\mse) & \plbRMS \\
$\sqrt{\chi^2_\nu}$  & \plbChiNu \\
\enddata
\end{deluxetable}

\begin{figure}
\epsscale{1.15}
\plotone{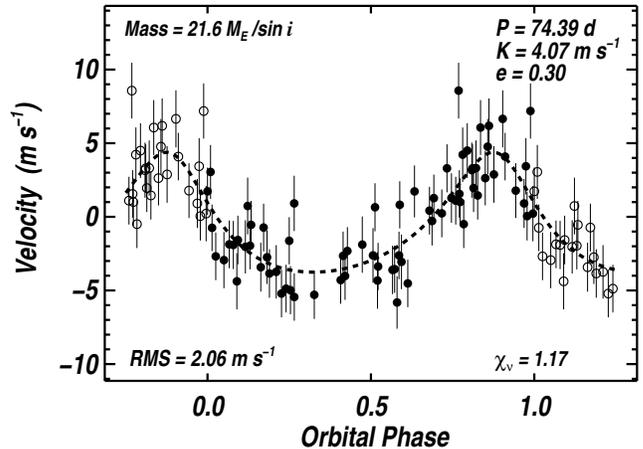}
\caption{Single-planet model for the RVs of Gl\,785, 
       as measured by Keck-HIRES.
       The dashed line shows the best-fit eccentric orbital solution.
       Symbols have the same meanings as in Figure \ref{fig:phased_97658}.
       }
\label{fig:phased_192310}
\end{figure}

We considered the null hypothesis for the observed periodic signal in the 
measured RVs of Gl\,785 by computing an FAP using the method described 
in Section \ref{sec:hd97658}, including allowing for eccentric solutions with the scrambled data sets.  
We found that none of the $10^3$ scrambled data sets had a larger  $\Delta\chi^2$ 
than the measured velocities, implying an FAP of less than $\sim$0.001. 

With an rms of \plbRMS\,\ms and a featureless periodogram of velocity residuals to the one-planet model, 
we do not see evidence for a second detectable planet orbiting Gl\,785.

\section{Photometric Observations}
\label{sec:photometry}

We also acquired photometric observations of HD\,97658 with the T12 0.80~m 
automatic photometric telescope (APT), one of several automatic telescopes 
operated by Tennessee State University (TSU) at Fairborn Observatory 
\citep{Eaton2003}.  
Gl\,785 is too far South for this observatory. 
The APTs can detect short-term, low-amplitude brightness 
changes in solar-type stars resulting from rotational modulation in the 
visibility of active regions, such as starspots and plages 
\citep[e.g.,][]{Henry1995} and can also detect longer-term variations 
produced by the growth and decay of individual active regions and the 
occurrence of stellar magnetic cycles \citep[e.g.,][]{Henry1995b,Hall2009}.  
The TSU APTs can disprove the hypothesis that RV variations
are caused by stellar activity, rather than planetary reflex motion
\citep[e.g.,][]{Henry2000}.  Several cases of apparent periodic RV  
variations in solar-type stars induced by the presence of photospheric 
starspots have been discussed by \citet{Queloz2001} and \citet{Paulson2004}.  
Photometry of planetary candidate host stars is also useful to search for 
transits of the planetary companions \citep[e.g.,][]{Henry2000b,Sato05}.

The T12 0.80~m APT is equipped with a two-channel photometer that uses two 
EMI 9124QB bi-alkali photomultiplier tubes (PMTs) to make simultaneous 
measurements of a star in the Str\"omgren $b$ and $y$ passbands.  The T12 
APT is functionally identical to the T8 APT described in \citet{Henry1999}. 
The final data products are differential magnitudes in the standard Str\"omgren
system.

During the three consecutive observing seasons between 2008 January and 
2010 June, the APT acquired 318 differential brightness measurements of 
HD\,97658 with respect to the comparison star HD\,99518 ($V=7.71$, $B-V=0.343$, 
F0).  We combined the $b$ and $y$ differential magnitudes into
$(b+y)/2$ measurements achieving typical single measurement precision of 
1.5--2.0\,mmag \citep{Henry1999}.

\begin{figure}
\epsscale{1.1}
\plotone{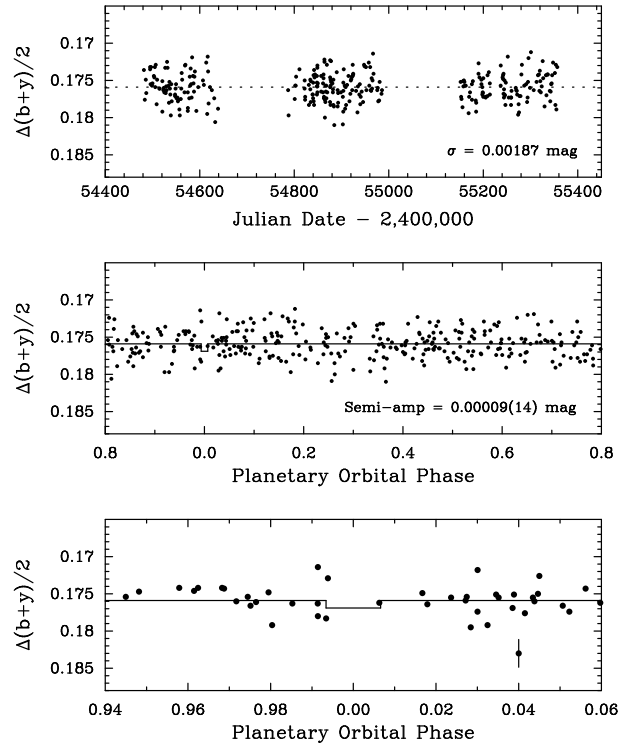}
\caption{Top panel:  The 318 Stro\"mgren $(b+y)/2$ differential 
magnitudes of HD\,97658 plotted against heliocentric Julian Date.  The 
standard deviation of these (normalized) observations from their mean 
(dotted line) is 1.99\,mmag.  Middle panel:  The observations plotted 
modulo the RV period.  Phase 0.0 corresponds to the 
predicted time of mid-transit.  A least-squares sine fit at the orbital 
period yields a semi-amplitude of  $0.09\pm0.14$\,mmag.  Bottom 
panel:  The observations near phase 0.0 plotted on an expanded scale.  The 
duration of a central transit is just three hours ($\pm0.0066$ phase units); 
the uncertainty of the transit time is $\pm0.64$ days ($\pm0.067$ phase
units).  The precision and phase coverage of our photometry are 
insufficient to determine whether or not shallow transits occur.}
\label{fig:photometry}
\end{figure}

The 318 measurements of HD\,97658 are plotted in the top 
panel of Figure \ref{fig:photometry}.  The second and third observing seasons have been 
normalized to match the mean brightness of the first 
season; the second and third year corrections were 1.75 and 0.70\,mmag, 
respectively.  This removes small year-to-year brightness changes in  
HD\,97658 and its comparison star and maximizes sensitivity to 
brightness variability on night-to-night timescales.  The standard deviation 
of the resulting normalized three-year data set is 1.87\,mmag, 
consistent with measurement error.  Periodogram analysis confirms the 
absence of periodic variability between one and 100 days. 

In the second panel of Figure \ref{fig:photometry}, the differential magnitudes are plotted
modulo the RV period.  Phase 0.0 corresponds to
the predicted time of mid-transit (Table~\ref{tab:orbital_params_hd97658}).  
A least-squares sine fit gives a semi-amplitude of $0.09 \pm 0.14$\,mmag. 
This tight limit to photometric variability at the RV period 
supports the hypothesis that the period RV signal is due stellar reflex motion 
from a planet in motion.

The observations near phase 0.0 are replotted on an expanded scale in the 
bottom panel of Figure \ref{fig:photometry}. The solid curve in the two lower panels 
approximates the depth (0.001 mag) and duration (three hours) of a central 
transit, derived from the orbital elements and assuming a water/ice 
composition for the planet.  The uncertainty in the time of mid-transit 
is approximately the width of the bottom panel.  The vertical error 
bar in the lower right of the transit window corresponds to the 
$\pm1.87$\,mmag measurement uncertainty of a single observation.  
The precision and phase coverage of our photometry are 
insufficient to detect shallow transits.

\section{Discussion}
\label{sec:discussion}

We announce two low-mass planets 
that were reported as anonymous ``planet candidates'' in \cite{Howard2010b}. 
HD\,97658\,b is a super-Earth planet with minimum mass 
$M_P\sin i$\,=\,\plaMsini\,$M_\earth$ in a $P$\,=\,\plaPer\,d orbit around a K1 dwarf star. 
Gl\,785\,b is a Neptune-mass planet with minimum mass 
$M_P\sin i$\,=\,\plbMsini\,$M_\earth$ in a $P$\,=\,\plbPer\,d orbit also orbiting a K1 dwarf. 

We see no evidence for transits of HD\,97658\,b, although our ephemeris and photometric phase coverage 
preclude detection of all but the deepest transits of a bloated planet.    
However, given the \textit{a priori} transit probability of 4\%, 
it is instructive to speculate about the transit signatures of various possible planet compositions.  
Using the models in \citet{Seager2007}, an 8\,\mearth planet composed of pure Fe, MgSiO$_3$, H$_2$O, or H 
would have radii $R_{\mathrm{pl}}$\,=\,1.3, 1.9, 2.4, and 5.5\,\rearthe, 
producing transits of depth 0.3, 0.6, 1.0, and 5.2\,mmag, respectively.  
These homogeneous planet models are oversimplified, but set the scale for admixtures of those ingredients.   
Transits of planets made of solids and water would have depths of $\sim$0.3--1.0\,mmag, 
while transits of a planet with a significant atmosphere could be much deeper.  

We have no constraints on transits of Gl\,785\,b because the host star is too far South for APT observations.  
The \textit{a priori} transit probability is 1\%.  
For comparison, we considered the transiting planets HAT-P-11b \citep{Bakos09} and 
HAT-P-26b \citep{Hartman2010}, which have masses 26\,\mearth and 19\,\mearth 
and radii 4.7\,\rearth and 6.3\,\rearthe, respectively.   
The implied densities, 1.38 and 0.42\,g\,cm$^{-3}$, suggest that these planets have considerable gas fractions.   
If GL\,785\,b has a radius in the range 4.7--6.3\,\rearthe, equatorial transits will be 4.4--7.8\,mmag deep.  
Such transits would be readily detectable from the ground, but would require a considerable observational 
campaign given the transit time uncertainty of $\pm$2.0\,d.

\citet{Fischer2005} showed that the occurrence of giant planets
with $K > 30$\,\ms correlates strongly with [Fe/H]. 
This has been interpreted as support for core accretion
models of exoplanet formation. However, low metallicity stars
might still be able to form less massive planets. 
\citet{Valenti10} noted that stars known to host \textit{only} 
planets less massive than Neptune (17\,\mearthe) tend to be metal poor relative to the Sun.  
HD\,97658 ([Fe/H] = \plaFeh, \msini = \plaMsini\,\mearthe) and 
Gl\,785 ([Fe/H] = \plbFeh, \msini = \plbMsini\,\mearthe) are consistent with this tentative threshold.  
Before interpreting this physically it is necessary to
check for metallicity bias in the subsample of stars around
which sub-Neptune mass planets can be detected with current techniques.  
Further, firmly establishing the apparent anti-correlation between 
host star metallicity and sub-Neptune mass planet occurrence
is best done with a well-controlled sample with uniform detection characteristics,
similar to \citet{Fischer2005}, or with well-understood detectability, similar to the Eta-Earth Survey. 

\acknowledgments{We thank the many observers who contributed to the measurements reported here.  
We gratefully acknowledge the efforts and dedication of the Keck Observatory staff, 
especially Scott Dahm, Hien Tran, and Grant Hill for support of HIRES 
and Greg Wirth for support of remote observing.  
We are grateful to the time assignment committees of the University of California, NASA, and NOAO  
for their generous allocations of observing time.  
Without their long-term commitment to RV monitoring, 
these long-period planets would likely remain unknown.  
We acknowledge R.\ Paul Butler and S.\ S.\ Vogt for many years
of contributing to the data presented here.
G.\,W.\,M.\ acknowledges NASA grant NNX06AH52G.  
G.\,W.\,H.\ acknowledges support from NASA, NSF, Tennessee State University, and
the State of Tennessee through its Centers of Excellence program.
This work made use of the SIMBAD database (operated at CDS, Strasbourg, France), 
NASA's Astrophysics Data System Bibliographic Services, 
and the NASA Star and Exoplanet Database (NStED).
Finally, the authors wish to extend special thanks to those of Hawai`ian ancestry 
on whose sacred mountain of Mauna Kea we are privileged to be guests.  
Without their generous hospitality, the Keck observations presented herein
would not have been possible.}


\LongTables  
\begin{deluxetable}{cccc}
\tabletypesize{\footnotesize}
\tablecaption{Radial Velocities and $S_{\mathrm{HK}}$ values for HD\,97658
\label{tab:keck_vels_hd97658}}
\tablewidth{0pt}
\tablehead{
\colhead{}         & \colhead{Radial Velocity}     & \colhead{Uncertainty}  & \colhead{}  \\
\colhead{JD -- 2440000}   & \colhead{(\mse)}  & \colhead{(\mse)}  & \colhead{$S_\mathrm{HK}$}
}
\startdata
 13398.04143 &    3.40 &    0.78  &          0.197                     \\ 
 13748.03543 &    1.41 &    0.79  &          0.190                     \\ 
 13806.96152 &    3.56 &    0.79  &          0.187                     \\ 
 14085.15873 &   -2.56 &    0.83  &          0.178                     \\ 
 14246.87902 &   -3.08 &    0.73  &          0.176                     \\ 
 14247.83980 &   -5.21 &    1.07  &          0.175                     \\ 
 14248.94470 &   -0.60 &    1.16  &          0.169                     \\ 
 14249.80244 &    1.19 &    1.24  &          0.174                     \\ 
 14250.83983 &   -0.72 &    0.99  &          0.174                     \\ 
 14251.89455 &    0.64 &    1.09  &          0.172                     \\ 
 14255.87104 &   -1.07 &    0.79  &          0.174                     \\ 
 14277.81740 &   -1.94 &    1.04  &          0.177                     \\ 
 14278.83838 &    1.33 &    1.03  &          0.175                     \\ 
 14279.83000 &    1.07 &    1.00  &          0.176                     \\ 
 14294.76351 &   -0.07 &    1.15  &          0.169                     \\ 
 14300.74175 &    3.77 &    1.22  &          0.172                     \\ 
 14304.76223 &   -2.53 &    1.23  &          0.174                     \\ 
 14305.75910 &   -0.11 &    0.81  &          0.174                     \\ 
 14306.77175 &    3.55 &    1.15  &          0.169                     \\ 
 14307.74725 &    4.39 &    0.83  &          0.175                     \\ 
 14308.75077 &    6.43 &    0.84  &          0.176                     \\ 
 14309.74773 &    5.28 &    1.23  &          0.176                     \\ 
 14310.74343 &    4.32 &    1.20  &          0.175                     \\ 
 14311.74391 &    7.30 &    1.15  &          0.176                     \\ 
 14312.74242 &   -0.26 &    1.18  &          0.177                     \\ 
 14313.74419 &   -1.57 &    1.26  &          0.178                     \\ 
 14314.75074 &    2.20 &    1.22  &          0.174                     \\ 
 14455.15432 &   -5.71 &    1.18  &          0.182                     \\ 
 14635.79759 &   -1.70 &    1.09  &          0.175                     \\ 
 14780.12544 &   -4.63 &    1.22  &          0.177                     \\ 
 14807.09051 &   -2.07 &    1.26  &          0.173                     \\ 
 14808.15781 &    2.09 &    1.30  &          0.171                     \\ 
 14809.14349 &    2.48 &    1.16  &          0.173                     \\ 
 14810.02507 &    8.16 &    1.29  &          0.173                     \\ 
 14811.11469 &    2.77 &    1.28  &          0.173                     \\ 
 14847.11818 &   -0.50 &    1.40  &          0.172                     \\ 
 14927.89832 &    3.45 &    1.37  &          0.170                     \\ 
 14928.96319 &   -2.78 &    1.30  &          0.170                     \\ 
 14929.84171 &   -3.59 &    1.22  &          0.169                     \\ 
 14954.97010 &    1.71 &    1.13  &          0.171                     \\ 
 14955.92258 &    2.61 &    0.59  &          0.172                     \\ 
 14956.90564 &    3.79 &    0.64  &          0.172                     \\ 
 14963.96612 &    4.04 &    0.66  &          0.169                     \\ 
 14983.87266 &    0.73 &    0.70  &          0.170                     \\ 
 14984.90278 &   -0.75 &    0.71  &          0.171                     \\ 
 14985.84542 &   -2.55 &    0.69  &          0.171                     \\ 
 14986.88960 &   -3.00 &    0.69  &          0.170                     \\ 
 14987.89549 &   -4.46 &    0.68  &          0.170                     \\ 
 14988.84400 &   -4.65 &    0.66  &          0.170                     \\ 
 15041.75244 &    7.08 &    1.35  &          0.169                     \\ 
 15164.11579 &    4.71 &    1.31  &          0.173                     \\ 
 15188.15802 &   -0.91 &    0.76  &          0.170                     \\ 
 15190.13283 &   -4.78 &    0.71  &          0.170                     \\ 
 15191.16082 &   -1.50 &    0.77  &          0.170                     \\ 
 15192.12820 &    1.97 &    0.69  &          0.171                     \\ 
 15193.11592 &    3.52 &    0.67  &          0.172                     \\ 
 15197.14316 &   -0.24 &    0.71  &          0.171                     \\ 
 15198.06394 &   -1.62 &    0.73  &          0.172                     \\ 
 15199.08955 &   -2.12 &    0.72  &          0.172                     \\ 
 15256.95777 &    3.84 &    0.71  &          0.180                     \\ 
 15285.94217 &   -1.43 &    0.68  &          0.175                     \\ 
 15289.83015 &    0.99 &    0.64  &          0.178                     \\ 
 15311.78396 &   -4.52 &    0.66  &          0.173                     \\ 
 15312.85958 &   -2.93 &    0.62  &          0.173                     \\ 
 15313.76751 &    0.82 &    0.65  &          0.172                     \\ 
 15314.78094 &    3.73 &    0.65  &          0.172                     \\ 
 15317.96407 &    1.70 &    0.65  &          0.174                     \\ 
 15318.94543 &   -3.46 &    0.67  &          0.175                     \\ 
 15319.90113 &   -4.34 &    0.66  &          0.176                     \\ 
 15320.85915 &   -5.55 &    0.57  &          0.180                     \\ 
 15321.83386 &   -2.68 &    0.62  &          0.181                     \\ 
 15342.87812 &   -1.40 &    0.63  &          0.176                     \\ 
 15343.82903 &   -1.37 &    0.67  &          0.176                     \\ 
 15344.88076 &    0.84 &    0.73  &          0.175                     \\ 
 15350.78135 &   -4.25 &    0.62  &          0.173                     \\ 
 15351.88526 &   -0.04 &    0.63  &          0.174                     \\ 
 15372.75655 &    2.37 &    0.63  &          0.179                     \\ 
 15373.78353 &   -0.22 &    0.60  &          0.179                     \\ 
 15374.75786 &   -0.32 &    0.61  &          0.178                     \\ 
 15375.77512 &   -1.73 &    0.61  &          0.177                     \\ 
 15376.74467 &   -1.66 &    0.60  &          0.177                     \\ 
 15377.74062 &   -0.77 &    0.59  &          0.177                     \\ 
 15378.74257 &    3.55 &    0.65  &          0.176                     \\ 
 15379.79041 &    0.84 &    0.63  &          0.176                     \\ 
 15380.74378 &    6.24 &    0.60  &          0.175                     \\ 
 15400.74241 &    1.31 &    0.72  &          0.177                     \\ 
 15401.76937 &    2.23 &    1.41  &          0.181                     \\ 
 15403.73903 &   -1.12 &    0.74  &          0.176                     \\ 
 15404.73645 &   -3.00 &    0.67  &          0.181                     \\ 
 15405.74110 &   -3.61 &    0.69  &          0.181                     \\ 
 15406.73695 &   -1.93 &    0.68  &          0.182                     \\ 
 15407.75726 &    2.44 &    0.81  &          0.180                     \\ 
 15410.73803 &    3.93 &    0.67  &          0.179                     \\ 
 15411.73488 &    0.95 &    0.71  &          0.178                     \\ 
 15412.73197 &   -0.23 &    1.26  &          0.178                     \\ 
 15413.73512 &    4.40 &    0.74  &          0.163                     \\ 
\enddata
\end{deluxetable}

\begin{deluxetable}{cccc}
\tabletypesize{\footnotesize}
\tablecaption{Radial Velocities and $S_{\mathrm{HK}}$ values for Gl\,785
\label{tab:keck_vels_gj785}}
\tablewidth{0pt}
\tablehead{
\colhead{}         & \colhead{Radial Velocity}     & \colhead{Uncertainty}  & \colhead{}  \\
\colhead{JD -- 2440000}   & \colhead{(\mse)}  & \colhead{(\mse)}  & \colhead{$S_\mathrm{HK}$}
}
\startdata
 13237.92941 &    1.73 &    0.59  &         0.2103                     \\ 
 13301.71519 &    4.76 &    1.13  &         0.2260                     \\ 
 13549.02705 &   -2.75 &    1.02  &         0.2040                     \\ 
 13926.01730 &   -1.63 &    0.56  &         0.2023                     \\ 
 13982.83072 &   -0.75 &    0.50  &         0.1963                     \\ 
 14247.08230 &   -3.60 &    0.67  &         0.1955                     \\ 
 14248.11326 &   -5.82 &    0.96  &         0.1950                     \\ 
 14249.12216 &   -3.06 &    1.09  &         0.1920                     \\ 
 14252.08848 &    1.73 &    0.93  &         0.1880                     \\ 
 14256.08153 &   -0.29 &    0.70  &         0.1890                     \\ 
 14279.03644 &    0.22 &    1.14  &         0.1920                     \\ 
 14280.04184 &    3.05 &    1.05  &         0.1940                     \\ 
 14286.03340 &   -4.38 &    1.18  &         0.1910                     \\ 
 14294.99649 &   -3.74 &    1.02  &         0.1950                     \\ 
 14634.06380 &    8.57 &    1.15  &         0.1790                     \\ 
 14634.98879 &    4.22 &    1.08  &         0.1830                     \\ 
 14636.03115 &    4.50 &    1.10  &         0.1820                     \\ 
 14637.06862 &    3.17 &    1.12  &         0.1820                     \\ 
 14638.02072 &    3.31 &    1.13  &         0.1830                     \\ 
 14639.05307 &    6.06 &    1.11  &         0.1850                     \\ 
 14640.12929 &    2.63 &    1.08  &         0.1850                     \\ 
 14640.97219 &    6.18 &    1.10  &         0.1850                     \\ 
 14642.09539 &    2.88 &    1.20  &         0.1870                     \\ 
 14644.10213 &    6.65 &    1.23  &         0.1870                     \\ 
 14688.96417 &   -2.62 &    1.09  &         0.1820                     \\ 
 14689.98535 &   -4.33 &    1.20  &         0.1830                     \\ 
 14723.77286 &    3.43 &    1.19  &         0.1840                     \\ 
 14724.80700 &    7.18 &    1.12  &         0.1830                     \\ 
 14808.68992 &   -2.04 &    1.01  &         0.1820                     \\ 
 14984.07717 &   -1.89 &    1.19  &         0.1750                     \\ 
 15019.01660 &    1.78 &    1.10  &         0.1740                     \\ 
 15026.96895 &   -2.94 &    1.18  &         0.1740                     \\ 
 15042.96436 &    0.91 &    1.13  &         0.1750                     \\ 
 15073.75665 &    0.42 &    0.61  &         0.1760                     \\ 
 15074.75183 &    1.27 &    0.55  &         0.1797                     \\ 
 15077.74110 &    3.29 &    0.64  &         0.1780                     \\ 
 15078.76189 &    1.28 &    0.76  &         0.1770                     \\ 
 15079.73545 &    1.10 &    0.62  &         0.1770                     \\ 
 15080.73918 &    1.01 &    0.58  &         0.1777                     \\ 
 15084.72917 &    1.45 &    0.65  &         0.1743                     \\ 
 15106.75946 &    0.73 &    1.21  &         0.1720                     \\ 
 15109.74590 &   -3.42 &    0.71  &         0.1740                     \\ 
 15111.71917 &   -3.85 &    0.68  &         0.1753                     \\ 
 15135.74754 &    0.65 &    0.64  &         0.1717                     \\ 
 15169.68272 &    0.91 &    0.72  &         0.1710                     \\ 
 15290.15433 &    0.82 &    0.56  &         0.1687                     \\ 
 15314.13774 &    4.09 &    0.59  &         0.1710                     \\ 
 15319.14050 &    0.05 &    0.70  &         0.1667                     \\ 
 15345.08584 &   -5.30 &    0.65  &         0.1720                     \\ 
 15351.09865 &   -4.30 &    0.58  &         0.1720                     \\ 
 15352.09190 &   -4.02 &    0.61  &         0.1717                     \\ 
 15374.11656 &    0.23 &    0.62  &         0.1743                     \\ 
 15378.11262 &    1.56 &    0.59  &         0.1740                     \\ 
 15379.10643 &   -0.50 &    0.62  &         0.1697                     \\ 
 15381.09845 &    3.29 &    0.63  &         0.1717                     \\ 
 15397.04238 &   -2.69 &    0.60  &         0.1720                     \\ 
 15400.11504 &   -1.87 &    0.61  &         0.1725                     \\ 
 15401.04500 &   -1.91 &    0.65  &         0.1717                     \\ 
 15402.08245 &   -1.58 &    0.69  &         0.1710                     \\ 
 15404.84477 &   -1.96 &    0.58  &         0.1727                     \\ 
 15405.08736 &   -0.55 &    0.64  &         0.1710                     \\ 
 15407.93295 &   -0.73 &    0.61  &         0.1753                     \\ 
 15412.01241 &   -5.21 &    0.60  &         0.1730                     \\ 
 15413.05434 &   -4.88 &    0.61  &         0.1740                     \\ 
 15414.04948 &   -4.98 &    0.65  &         0.1717                     \\ 
 15414.92114 &   -5.45 &    0.58  &         0.1740                     \\ 
 15426.03531 &   -2.65 &    0.68  &         0.1717                     \\ 
 15427.00892 &   -2.32 &    0.62  &         0.1710                     \\ 
 15433.99704 &   -3.37 &    0.58  &         0.1737                     \\ 
 15435.78071 &   -2.41 &    0.63  &         \nodata                    \\ 
 15436.75896 &   -2.30 &    0.63  &         \nodata                    \\ 
 15437.76291 &   -3.55 &    0.57  &         0.1747                     \\ 
 15438.76140 &   -2.62 &    0.59  &         0.1770                     \\ 
 15440.75917 &   -4.52 &    0.59  &         0.1773                     \\ 
 15455.73811 &    1.95 &    0.63  &         0.1750                     \\ 

\enddata
\end{deluxetable}

\bibliographystyle{apj}
\bibliography{eta3}

\enddocument